\documentclass[epj]{svjour}
\usepackage{graphicx}
\renewcommand{\v}[1]{\underline{#1}}

\begin{document}
\title{Hierarchy of Scales in Language Dynamics}
\author{Richard A Blythe
}                     
%
%
\institute{SUPA, School of Physics and Astronomy, University of Edinburgh, Peter Guthrie Tait Road, Edinburgh EH9 3FD, UK}
\date{Received: date / Revised version: date}
%
\abstract{
Methods and insights from statistical physics are finding an increasing variety of applications where one seeks to understand the emergent properties of a complex interacting system. One such area concerns the dynamics of language at a variety of levels of description, from the behaviour of individual agents learning simple artificial languages from each other, up to changes in the structure of languages shared by large groups of speakers over historical timescales. In this Colloquium, we survey a hierarchy of scales at which language and linguistic behaviour can be described, along with the main progress in understanding that has been made at each of them---much of which has come from the statistical physics community. We argue that future developments may arise by linking the different levels of the hierarchy together in a more coherent fashion, in particular where this allows more effective use of rich empirical data sets.
} 
\maketitle

\section{Introduction}
\label{intro}

In 1972, Anderson famously articulated the idea that ``more is different'' \cite{and72}: that the emergent consequence of fundamental physical laws at one scale lead to new physical laws at a larger scale, and ones that are as fundamental as those that apply at the smaller scale. In physics, statistical mechanics provides the conceptual and methodological framework to build the bridge between the dynamics at the larger scale and the interactions between the constituents at the smaller scale.

A good illustration of this procedure is provided by the two steps of coarse-graining that sit between the classical Hamiltonian dynamics of point particles and the Navier-Stokes equations for fluids. The first step is to shift from a description in terms of individual particles to one in terms of \emph{classes} of particles, for example, all those at a similar point in space and with a similar velocity (see e.g.~\cite{rei09}). This trick, pioneered by Maxwell to determine the equilibrium distribution of particle velocities \cite{max60}, reduces the dimensionality of the problem by a factor of $N$, the number of particles in the system. Since $N$ is extremely large (order $10^{23}$ or more) for macroscopic systems, this is an impressive reduction in complexity. However it comes at a cost: the full dynamics take the form of an infinite hierarchy of equations, each of which requires a solution for all of the others in order to be solved. To break this vicious circle, a physical insight is needed. Specifically, the molecular chaos assumption (employed by both Maxwell \cite{max60} and Boltzmann \cite{bol72} with varying degrees of explicitness) amounts to a statement that the details of collisions between particles is less important than their overall effect, which is taken to eliminate correlations between pairs of particles. The Boltzmann equation that results is a nonlinear equation that describes the population dynamics of particles and their movements between different classes. A further step of averaging yields coarse-grained density and velocity fields. Again, this step generates a hierarchy of equations which can be truncated by appealing to a physical insight, namely that the macroscopic fields of interest vary slowly on the length- and timescales of the microscopic particle dynamics \cite{gra63}. We thus arrive at the Navier-Stokes equation, which is highly nonlinear and can describe extremely rich physical states, such as turbulence, whose existence is unimaginable in the original Hamiltonian formulation.

The fact that Hamiltonian equations of motion lie at the bottom of this hierarchy is incidental to the statistical mechanical coarse-graining procedure. The crucial component is the physical insight that allows certain aspects of the dynamics at one level of the hierarchy to be replaced with an effective description that makes progress towards understanding the next level possible. There is no reason that such insights should be the preserve of condensed matter physics. Indeed, Anderson's hierarchy extended from particle physics to the social sciences, passing through cell biology and psychology on the way. It should therefore come as no surprise---nor concern---that the reach of statistical physics has profitably expanded into the biological \cite{vla11,cat12} and sociological \cite{cas09} domains. Furthermore, in line with Anderson's position, exploring these applications areas often raises new fundamental questions in nonequilibrium statistical mechanics.

One area that has received growing attention from statistical physicists over the past fifteen years or so is the quest for a  quantitative understanding of language dynamics. Human language exhibits complexity at multiple scales in at least two dimensions. First,  language itself exhibits complex structure at the level of sounds, words and sentences through \emph{phonology} (which specifies how sounds may be be combined), \emph{morphology} (how meaningful combinations of sounds may be further combined to make words) and \emph{syntax} (how sentences are formed) \cite{ogr11}.  At each level, the specific set of valid combinations varies from one language to the next: for example, the order with which words with a specific function appear in a sentence is not the same in every language (as we will discuss further in Section~\ref{sec:lineages} below). At the same time, variation is not completely arbitrary: for example, some word orders are more common across the world's languages than others. Meanwhile, each language changes over time at all levels of organisation, leading on occasion to predictable phenomena like vowel shifts \cite{lab94} and the unidirectional character of grammaticalisation \cite{hop03}, whereby words with more concrete meanings (like objects and actions, represented by nouns and verbs) take on more abstract grammatical functions (like relationships between objects, represented by prepositions) over time, whilst the reverse process is extremely rare \cite{has99}.

It is natural to assume that the process of language change is related to the variation that is observed between human languages. This leads to the second dimension of linguistic complexity, in that a language is a collective phenomenon, shared by a group of speakers, and that originates from interactions between individual members of this group (whose membership may also change over time).  It is this dimension that has been of particular interest to statistical physicists, since it concerns the question of how the properties of the system at large can be understood from the behaviour of the component parts.  This complex systems approach also has a long tradition in linguistics, its essence being captured by the \emph{usage-based model} \cite{lan88u} and leading to a view of language as a complex adaptive system articulated by various authors \cite{ste00,bec09}.

In this short Colloquium I focus more or less exclusively on this second social dimension, and in particular on the processes by which languages change and thereby differentiate themselves from one another over time, whilst retaining certain features in common. Empirical, experimental and modelling work in this area typically employs rather simple representations of language, for example, competition between two competing words with the same meaning, thereby abstracting away much of the structural complexity described above. However it is important to keep in mind that this structural complexity exists, and we will briefly return to the question of how to build on the framework set out in this Colloquium to incorporate more sophisticated language structure (and why this would be worthwhile) in the concluding section. 

Within this perspective of language as a dynamical object undergoing a process of cultural evolution through interactions between individual speakers endowed with certain cognitive and behavioural capabilities, we can identify four levels of a hierarchy that we discuss in separate sections of this article.  We begin in Section~\ref{sec:lineages} with the largest length- and timescales, where language is viewed as a characterising a group of speakers and undergoes change in its own right without direct reference to the behaviour of its users. In Section~\ref{sec:communities}, individual speakers enter the picture as agents adopting one (or more) of a number of pre-existing languages, thereby causing the prevalence of different languages to change over time. Section~\ref{sec:individuals} shifts focus from language as a whole to individual utterances, and in particular the process by which an innovative feature initially used by a small number of speakers can propagate through the entire community of language users, precipitating a change in its macroscopic state. Finally, in Section~\ref{sec:acts} we survey (mostly experimental) work that seeks to establish the cognitive and behavioural biases that underpins the linguistic behaviour of individual speakers and that are ultimately responsible for the processes that occur at higher levels of the hierarchy.

By taking a view of language as a primarily cultural evolutionary phenomenon, emerging in systems of agents with the necessary linguistic capabilities, the question of how these capabilities arise in the first place is largely excluded from our discussion. This is nevertheless a large, active and highly multi-disciplinary field of enquiry in its own right, and the interested reader is referred elsewhere \cite{chr03,fit10,hur14} for an overview of research in this area. We add however the cautionary note that the field of language origins is not without its controversies, particularly on the question on what biological adaptations have occurred---if any---that are specific to language (see e.g.~\cite{hau02,eva09} for contrasting views on this topic). Nevertheless, the hierarchy we have set out here could in principle be extended to longer timescales and to the coevolution of cultural and biological evolution, thereby potentially shedding light on some of these issues. We touch on such possibilities in the concluding section.

Even within the restricted range of topics covered here, space precludes detailed discussion of all the relevant literature. The main focus is therefore on a relatively small number of studies, each of which exemplifies a particular paradigm within the field of language dynamics. Although the cited references should serve as an entry-point to the wider literature, references to relevant review articles that go into more depth on specific topic are provided at appropriate points.

\section{Variation in language structure}
\label{sec:lineages}

As noted above, languages vary in their structure at all levels of organisation. A useful resource for studying similarities and differences between languages is the World Atlas of Language Structure (WALS, \cite{wals}) which specifies up to 192 features (for example, the size of the consonant inventory or the number of genders) across over 1000 languages.  This database reinforces the findings of typological research \cite{cro03} that whilst considerable variation is evident in the structure of different languages, this variation is not unconstrained. An illustrative example is basic word order, which concerns the order that the subject (S), object (O) and verb (V) appear in a sentence. Although all six orderings would convey the same information, the two patterns that have the subject in the first position characterise around $90\%$ of the languages in WALS. This highly nonuniform distribution over the possible states of a language is referred to as a \emph{typological universal}. More sophisticated universals also exist \cite{cro03}, in particular \emph{implicational universals}, whereby the presence of one structure in a language tends to imply another structure: for example, if adjectives precede nouns, it is very likely (around 90\% probability) that demonstratives also precede nouns \cite{wals}.

At the largest length and timescales we can think of each language as a single coherent unit, and defined in terms of some set of $F$ features. For simplicity, let each of these features be discrete such that feature $f$ takes one of $m_f$ distinct values, denoted $\sigma_f$. For example, in the case of basic word order, $\sigma_f$ takes one of six values. At this level of description, the state of  a language is fully specified by the vector $\v{\sigma} = (\sigma_1, \sigma_2, \ldots, \sigma_F)$. The first type of typological universal then corresponds a nonuniform distribution over the set of possible states $\{\v{\sigma}\}$, while implicational universals correspond to correlations between components of the vector $\v{\sigma}$. In this framework, one can postulate (at least) two possible explanations for these typological universals. The first is that some states $\v{\sigma}$ are more stable than others: for example, languages that place the subject first are less likely to change than those that have it in a different position. The second is to acknowledge that (like species) languages may have common ancestors (for example, French and Romanian are descendants of Latin), and hence correlations between the states of different languages can be attributed to a common historical period.  The main method to discriminate such explanations is phylogenetic analysis (see e.g.~\cite{mac05} for a review in the context of cultural evolution). 

\begin{figure}
\begin{center}
\includegraphics[width=0.9\linewidth]{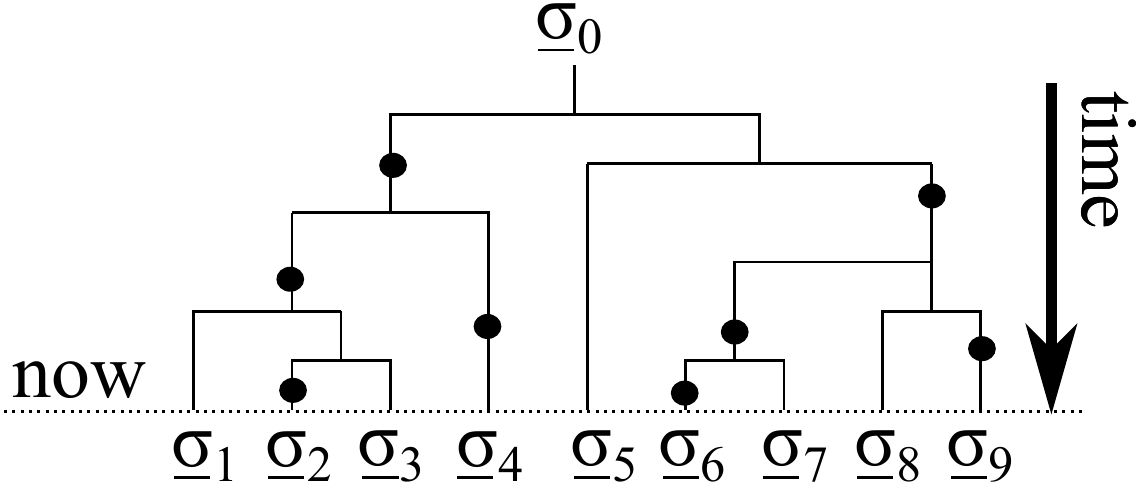}
\end{center}
\caption{\label{fig:tree} A tree relating languages with different structures. New languages are created when an ancestor splits into two daughters. Changes in state occur at points in time marked with a filled circle. This gives rise to a set of distinct languages, $\v{\sigma}_1, \ldots, \v{\sigma}_9$, at the present time that are all descended from the common language $\v{\sigma}_0$.}
\end{figure}

The basic idea is to assume first of all that the historical evolution of languages can be modelled as a tree within which branches occur at points in time where an ancestor language splits into two daughters---see Figure~\ref{fig:tree}. On top of this tree structure, a stochastic model for the changes in state that occur is superimposed. In linguistics, this type of model has been referred to as a \emph{state-process} model and is attributed to Greenberg \cite{gre78}. A natural choice for the stochastic dynamics is a time-homogeneous Poisson process wherein the probability that a language in state $\v{\sigma}$ changes to state $\v{\sigma}'$ in a time interval ${\rm d}t$ is
\begin{equation}
\label{poisson}
P_{{\rm d} t}(\v{\sigma}'|\v{\sigma};t) = \omega(\v{\sigma}'|\v{\sigma}) {\rm d} t \;,
\end{equation}
i.e., that the rate of change from one state to another is constant in time across all branches of the tree. Sampling from this distribution gives a set of points at which changes in state occurred, shown as circles on the branches in Figure~\ref{fig:tree}. From the stochastic model for changes in state, one can compute the likelihood of arriving at the current distribution of languages (given a common ancestor and specific tree structure). Then, using Bayesian statistics, one can infer a posterior distribution of the parameters in the model, which includes the tree structure and the transition rates between language states. It is also possible to incorporate further constraints where information is known, for example, points in history where a language split occurred.

Since sampling the full parameter space to obtain posterior distributions is computationally demanding, it is only recently that these methods have become widely-used. Prior to this, other methods---such as maximum parsimony methods---were used to infer the relationships between different languages \cite{gra00}. In this context, maximum parsimony typically corresponds to finding the shortest evolutionary pathways between different languages, for example, the sequence of events with the smallest number of changes in state over the entire history. Initial studies \cite{gra00,gra03} focussed on the lexicon (i.e., the set of words available to speakers to express meanings). The idea is the closely-related languages have more words in common than distantly-related languages. More formally, one can define \emph{cognate sets} that comprise words from different languages that are judged to have the same form for the same meaning \cite{ogr11}. In this context, each feature $f$ corresponds to one of the cognate sets, and $\sigma_f=0$ or $1$ depending on whether the language in question appears in cognate set $f$. A change in state  then corresponds to a language entering or leaving a cognate set.

One obvious question is whether it is reasonable to assume that relationships between languages form a tree-like structure, as in Figure~
\ref{fig:tree}. Evidence in favour of this assumption was provided by Gray \textit{et al} \cite{gra00}: the most parsimonious tree was more consistent with an `express-train' hypothesis where languages were established during a rapid process of colonisation of successive `stations' in the Pacific by Austronesian-speaking populations, as opposed to an `entangled bank' hypothesis characterised by a high degree of contact between the different languages.

More recent work has moved in two parallel directions. One is that described above, that is, to incorporate explicit stochastic models of language change processes, and to use Bayesian statistics as a means to estimate parameters that characterise the dynamics. The second direction is to employ more sophisticated measures of distance between two languages (such as the relative entropy \cite{ben02} or Levenshtein distance \cite{pom11} between two sets of words in different languages with the same meaning) and use variations on the maximum parsimony approach to determine likely tree structures. In principle, both approaches could be combined; however this is not yet widely practised, presumably because more sophisticated notions of language structure translate to correspondingly more complex stochastic dynamical models.

Taken together, these complementary investigations have delivered a wide variety of findings, a few of which we touch on here.  First, they provide support for the theory that Indo-European languages originated in Anatolia around 8,000--9,500 years ago \cite{gra03,bou12} and that inferences based on the structure of the sound system, word order and other grammatical features allows reconstruction to greater time depths than is possible with the lexicon, and specifically that Papuan languages diverged over 10,000 years ago \cite{dun05}. Using the dynamical approach, the rate of lexical change has been estimated to vary by up to two orders of magnitude, and to be correlated with the frequency of word use \cite{pag07} and population size \cite{bro15}. In both cases the data are suggestive of a power-law dependence of the rate on the explanatory variable. A power law relationship between the rate of verb regularisation and verb frequency was also found from a corpus of English texts \cite{lie07}. Here, however, the statistical physicist will likely be disappointed that the exponents of the power laws relating the rate of language change to the frequencies of different linguistic units are not the same: this suggests that the underlying mechanisms generating these mechanisms may not be universal---at least in a strict statistical physics sense.

Most relevant to questions about the origin of typological universals, and in particular the question of whether more common structures are also more stable, are two studies of word-order universals \cite{dun11,mau14}. Of these two studies, that of Maurits and Griffiths \cite{mau14} is conceptually more straightforward. Here, the authors find that despite SOV word order being slightly more common than SVO order in the contemporary distribution of languages (at 48\% and 41\% respectively \cite{wals}), the historical rates of change between the two orders are indistinguishable, suggesting that they are in fact of equal stability.  Dunn \textit{et al} \cite{dun11} meanwhile investigate the  more subtle question of implicational universals. Here, the hypothesis is that if the orders of two word classes (say adjective-noun and demonstrative-noun) are found to be correlated in the contemporary distribution, one would expect to see a similar correlation in the historical rates of change of these two structures. Only two of the correlations expected from the contemporary distribution were observed in the historical dynamics of more than one language family. Furthermore, most of the dynamical correlations that were identified were present in a single language family.

This latter result suggests that factors specific to individual cultures contribute to the dynamics that give rise to typological universals. However, at this level of description, it does not pinpoint what these factors might be. Some clues are provided by investigations of correlations between language structures and quantities that vary between cultures: of these, two studies \cite{ded07,lup10} have been particularly prominent---in part due to a degree of controversy. Dediu and Ladd \cite{ded07} presented evidence that two genes related to brain structure potentially contribute towards a cognitive bias that disfavours linguistic tone (the use of pitch to distinguish words) in the process of language acquisition. Meanwhile, Lupyan and Dale \cite{lup10} demonstrated an inverse relationship between language size (as measured number of speakers or geographical area) and structural complexity across 28 distinct features (e.g., the number of cases): in other words, that more widely-spoken languages tend to be less complex. Further clues might emerge by appealing to the dynamics of language at a lower level of description.

\section{Languages competing for speech communities}
\label{sec:communities}

The model of language in the previous section made explicit reference only to its structure (as defined by a set of features). Any influence of the underlying population of speakers was incorporated at best implicitly, for example, by allowing the transition rates between states to depend on culture-specific factors like its size (as was done in \cite{bro15}). In all of these studies, interactions between distinct languages, other than the possibility that they share an ancestor, were ignored. By descending a level into the hierarchy, we can investigate interactions between languages mediated by the groups of speakers using them. Note that in order to distinguish from any other social groups (e.g., populations of a country) we shall use the term \emph{speech community} (borrowed from linguistics) to describe a social group with a common language.

The obvious example of an interaction between two or more languages is where they are spoken in the same geographical area, and thereby compete for a common pool of speakers. The Celtic languages (Welsh, Scottish and Irish Gaelic, Cornish and Manx) that spoken alongside English within the British Isles provide an example of such a set \cite{bel08}. The increasing dominance of English over the Celtic languages---to the extent of extinction in the case of Cornish and Marx---is evidence of competition between them that is independent of the population dynamics in the relevant regions, since neither region suffered a mass population extinction in the relevant historical period. Rather it is the process of \emph{language shift}, whereby individual speakers switch from one language to another through a variety of mechanisms \cite{tho01}, that is believed to explain such declines.

A seminal model of language shift was proposed by Abrams and Strogatz \cite{abr03}. It is couched in terms of the fractions $x$ and $y=1-x$ of a population speaking two languages $\rm X$ and $\rm Y$ respectively. The fraction of speakers who shift from $\rm Y$ to $\rm X$ per unit time is denoted $P_{\rm YX}$ and is taken to depend on the size, $x$, of speech community of $\rm X$ and a quantity $0 \le s \le 1$ that encodes the `status' of $\rm X$. The status of $\rm Y$ is given implicitly by $1-s$, so for $s>\frac{1}{2}$, $\rm X$ has a higher status than $\rm Y$. Ignoring all fluctuations, one arrives at an ordinary differential equation for $x$ of the form
\begin{equation}
\label{as}
\frac{{\rm d}x}{{\rm d}t} = y P_{\rm YX}(x,s) - x P_{\rm XY}(x,s) \;.
\end{equation}
The transition rates $P_{\rm YX}(x,s)$ and $P_{\rm YX}(x,s)$ are not independent. If the size and status of a language are the only factors that affect the competition between them, we must obtain the same dynamics under relabelling of ${\rm X}$ and ${\rm Y}$. This implies that $P_{\rm XY}(x,s) = P_{\rm YX}(1-x,1-s)$. Under the further mild assumptions that (i) languages with no speakers are not spontaneously reinvented; (ii) that a language with zero status is never shifted towards; and (iii) that $P_{YX}(x,s)$ increases monotonically in both arguments, the fixed points of (\ref{as}) can be determined.  One finds that there are always two stable fixed points at $x=0$ and $x=1$, corresponding to extinction of ${\rm X}$ and ${\rm Y}$ respectively, and a third fixed point that corresponds to coexistence of both languages. This latter fixed point is always unstable, implying that one of the two languages is doomed to extinction. Which of the two languages that is destined for this fate depends on the initial condition.

Although extremely simple, this model provides a number of valuable insights. First, Abrams and Strogatz \cite{abr03} showed that the model can be fit to time-series data for a variety of languages. To achieve this, they chose the form of the transition rate to be $P_{\rm YX}(x,s) = c s x^a$, where $a$, $c$ and $s$ are a set of parameters fit separately to each linguistic time series (along with $x(0)$). The value of the exponent $a\approx1.3$ was reported to be fairly robust across the languages, although little was said about the remaining parameters. Nevertheless, this appears to allow confident prediction of the fate of a minority language, and in particular the timescale over which extinction may be expected. From a theoretical perspective, the most interesting parameter in this model is $s$, the social status of language $\rm X$. The thinking here is that speakers have some awareness of the opportunities afforded to them by their choice of language, for example, to improve their own personal wealth. At this level of description, however, the precise mechanism by which this awareness is gained is left unspecified.

The main value of Abrams and Strogatz' model is the limited range of outcomes it predicts under broad but reasonable assumptions. In particular, the fact that coexistence of two languages of different social status is impossible in this framework has motivated many extensions of the basic model, a number of which are reviewed in greater depth elsewhere, e.g.~\cite{kan09,sol10}. There are at least two ways in which two languages of unequal status can coexist. The first is to assume that initially the languages are spoken in different regions and only then mediate contact between them by diffusion of speakers of one language into the geographical region in which the other was originally spoken \cite{pat04}. This mechanism works because Eq.~(\ref{as}) has stable fixed points at $x=0$ and $x=1$, i.e., when either of the two languages is the only one being spoken. This stability implies a small incursion of speakers of the other language can be tolerated. What is interesting about this result is that demonstrates the possibility for a \emph{globally} disfavoured language (the one with the lower status) to survive alongside the high-status language. 

Is there any way in which two languages of different status might coexist at the same point in space?  This was shown to be possible if one combines the process of language shift in the Abrams-Strogatz model with a process of reproduction that causes both the population size to grow and for the offspring to inherit the language of its parents \cite{pin06}. In common with many other models of population growth, this model featured a carrying capacity which limits the maximum size of the population \cite{mur93}. However, in \cite{pin06} a separate carrying capacity was assigned to each speech community, meaning that when the region is saturated by speakers of language $\rm X$, it can nevertheless continue to accommodate speakers of language $\rm Y$. It turns out that it is this assumption that facilitates coexistence. If one instead assumes that the overall size of the \emph{combined} population, comprising both speech communities, is limited by a carrying capacity, which is reasonable if they compete for the same set of resources to sustain their population, the possibility of coexistence is removed \cite{kan08}. It is however possible to engineer coexistence if the relative social status of the two languages is different in different parts of space, which could plausibly arise because ultimately the status is a subjective judgement made by individual speakers, and consequently different groups of speakers could reach different judgements.  Taking the results from this body of research together, it seems that the usual outcome of language competition is for a dominant language to drive others to extinction, unless there is some spatial symmetry-breaking present, either in the initial condition or through variation of an external factor like social status. This generalisation also applies when bilingualism is also incorporated into the Abrams-Strogatz model \cite{min08}, in the presence of noise \cite{sta05} or both \cite{cas06}, although long-lived metastable states in which both languages persist can been seen. The main departure from this generalisation is found under external interventions that are explicitly designed to prevent extinction of declining languages, such as embedding them in the school curriculum \cite{min08}.

An empirical fact worth noting is that extinction of one of the languages in contact is not the only possible outcome. In certain instances, notably after colonisation by Europeans of regions inhabited by speakers of a different language, a new language, known as a creole, can form \cite{muf08}. A sister project to the World Atlas of Language Structures (WALS, \cite{wals}), known as the Atlas of Pidgin and Creole Language Structures (APiCS, \cite{apics}), documents the structural features of these languages formed through contact.  However, quantitative modelling of the formation process is as yet in its infancy. In this regard, a notable contribution \cite{tri15} involves a stochastic model with the flavour of the naming game (described in more detail in the next section) quantitatively reproduced the empirical finding that the European language tended to dominate when the fraction of Europeans in the population is high, whilst at lower levels of colonisation a creole tends to form.

\section{Emergence of a common language}
\label{sec:individuals}

In the discussion so far, a language has been defined at the speech community level, for example, as a set of structural features that are characteristic of the language as a whole, and help to differentiate it from languages spoken by different groups of people. Nevertheless, there is considerable variation in the way language is used by speakers of the same language. Indeed this variation exists both within and between individual members of a speech community \cite{ogr11,cro00,mey11}. In common with structural variation between languages, this variation exists at all levels of linguistic complexity: speakers can differ in the way they pronounce the same word or a syllable that appears in a class of words, in the word they use to convey a given meaning, or in they way they construct sentences. This variation may arise in a number of different ways \cite{cro00}. For example, one property of language is its open-endedness which confers on speakers the ability to express propositions that they have never articulated or heard before. This entails producing new words or recombining words in a new way. At the same time, speakers may adjust their set of grammatical rules, perhaps to achieve greater internal consistency of the system (e.g., by regularising an irregular verb). Alternatively, words and phrases may enter one language by borrowing from another (see, for example, the German word \emph{ansatz} that is widely used by physicists, and whose precise meaning is hard to convey efficiently in English).

These innovations explain why some unconventional forms may coexist alongside the conventional forms that characterise a language at the speech community level. However, these conventions also change over time, as we have seen in Section~\ref{sec:lineages}. An important question relates to the mechanism by which these innovations, which are created in interactions between small numbers of agents, are propagated and result in a change in a language's macroscopic state. By now, this problem of \emph{conventionalisation} or \emph{consensus-formation} is a very well-studied problem in statistical physics, with progress having been very comprehensively reviewed fairly recently \cite{cas09}. It is also worth noting that the field of sociolinguistics is devoted to understanding variability in language use, the factors that result in one variant form being used over another, and how these factors contribute to language change \cite{mey11}.

In order to discuss this large body of work in a consistent fashion, we shall adopt Croft's evolutionary framework for language change initially set out in \cite{cro00}. This identifies three distinct dynamical processes that contribute. The first is replication, which is the reproduction of structures (e.g., sounds, words or a particular ordering of word classes in a sentence). The second is innovation, which may take the form of a completely novel utterance or (more likely) some small change to a structure as it is replicated (e.g., putting words in a different order in order to suggest a novel meaning to the listener). The third process is selection, which is the preferential replication of one structure over another. Later works \cite{bax09,bly12} introduce a distinction between two types of selection: \emph{interactor selection}, whereby certain individual members of the speech community have a greater influence on a speaker's behaviour than others, and \emph{replicator selection}, whereby one structure is systematically replicated in preference to another by a speaker or a group of speakers.

\subsection{Paradigmatic models of consensus formation}

Different aspects of the process of consensus formation have resulted in models that combine innovation, replication and selection in a variety of different ways. We frame the discussion with two paradigmatic and well-studied models, namely the Moran model of replication dynamics, originally formulated in the context of population genetics but more recently co-opted for linguistic purposes, and the naming name, which was inspired by experiments in which robots perceive natural stimuli and attempt to communicate with one another \cite{ste03}.

\subsubsection{The Moran model of replication dynamics}
\label{sec:moran}
%
\begin{figure*}
\begin{center}
\includegraphics[width=0.6\linewidth]{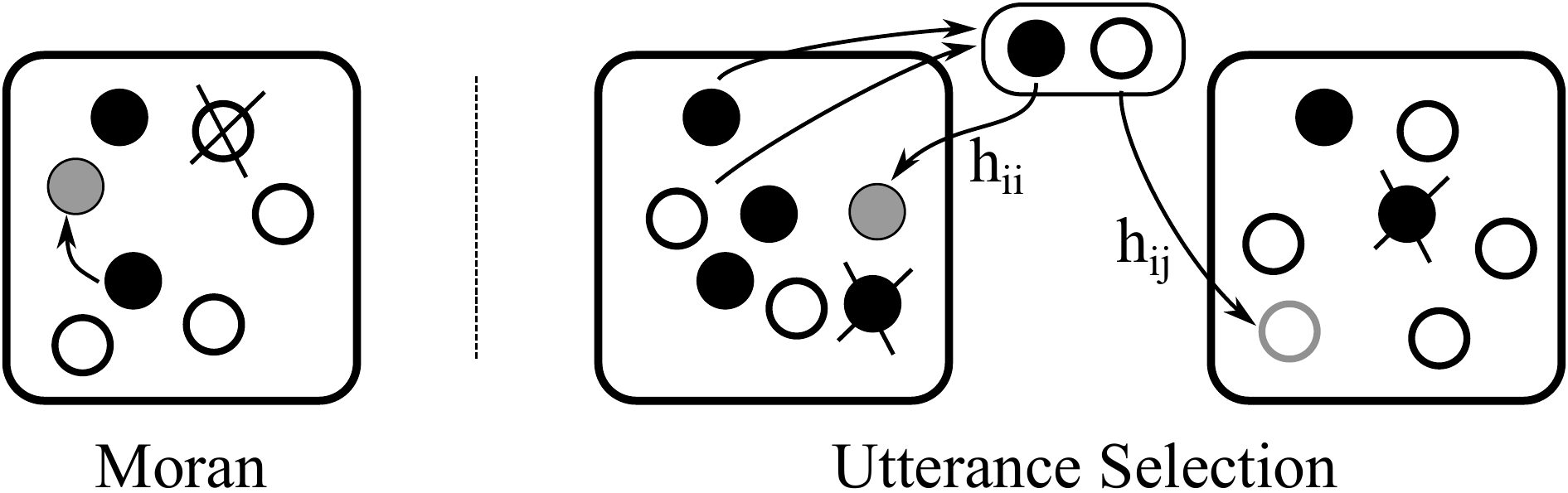}
\end{center}
\caption{\label{fig:moran} The Moran and utterance selection models of replication dynamics. In the Moran model, two agents are chosen at random from the speech community. One (shown crossed-out) has its state replaced with a copy of another (shown light shaded). The utterance selection model is defined in terms of units of linguistic variation, rather than agents. Each speech community now represents a memory of linguistic events. In the figure, the left-most store is that of a speaker (agent $i$), and the right-most store that of a listener (agent $j$). A set of (here two) tokens is sampled from the speaker's store, shown intermediate between the two stores. With probability $h_{ii}$, a copy of a produced token is placed in the speaker's store, displacing an existing memory in the process. With probability $h_{ij}$, a copy is stored in the listener's store. In the original formulation of the utterance selection model, both agents in an interacting pair act as speakers and listeners in each interaction. The dynamics (up to a factor of two in the characteristic timescale) are the same.}
\end{figure*}

The Moran model applies to a fixed-sized population of $N$ agents, each of whom can be in one of a number of discrete states. The simplest case has just two states, ${\rm A}$ and ${\rm B}$, which in linguistic applications relate to speakers who use different pronunciations of the same sound, different words for an object, or different grammatical rules with the same function (e.g., marking the past tense).  In each elementary update of the Moran model, two agents are chosen at random. One is designated as the speaker, and the other as the listener. The speaker produces an utterance that allows the listener to identify whether the speaker is a user of variant ${\rm A}$ or variant ${\rm B}$. As a result of the interaction, the listener adopts the same variant as that used by the speaker.  The update rule is illustrated in Figure~\ref{fig:moran}.

The Moran model serves as a simple mathematical description of accommodation, a process where interacting agents serve to align their linguistic behaviour \cite{gil79}. In this formulation it is implicit that each member of the speech community uses either ${\rm A}$ or ${\rm B}$ categorically, and not some variable mixture of the two, although the latter is in fact commonplace in linguistics \cite{mey11}. Also this model as formulated has no spatial or social structure: every pair of agents is equally likely to interact, and accommodation is equally likely in either direction. Extensions to the Moran model exist that incorporate variability in individual behaviour and in the interaction strengths between agents. One of these is the utterance selection model, also illustrated in Figure~\ref{fig:moran}. In this model, each individual speaker is represented by a population of stored tokens of linguistic behaviour. A pair of agents interacts with some probability $G_{ij}$, whereupon they produce a set of tokens by sampling from the store. This allows for individual speakers to exhibit variable linguistic behaviour. After the interaction, agent $i$ stores tokens produced by speaker $j$ in their store with a probability proportional to $h_{ij}$; agent $j$ does likewise with a probability proportional to $h_{ji}$. These two probabilities need not be symmetric: $h_{ij}\ne h_{ji}$, which then corresponds to the two speakers accommodating to a different degree.  In this way, interactor selection (social biases) can be incorporated. As we will see below, many insights from the Moran model can be used to understand the behaviour of more complex models like the utterance selection model.

\subsubsection{The naming game}

\begin{figure}
\begin{center}
\includegraphics[width=0.66\linewidth]{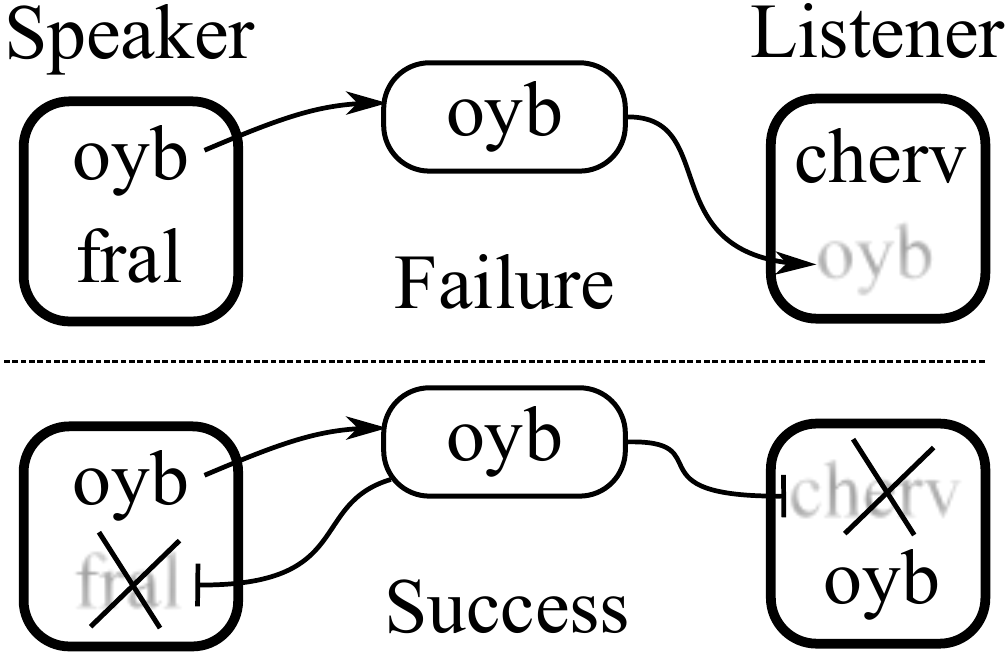}
\end{center}
\caption{\label{fig:namgam} The naming game. In this model, agents are equipped with an expandable inventory of names for a single object. When a pair of agents interact, the speaker samples a name uniformly from their inventory (inventing a word if it is empty). In both cases shown here, this inventory comprises the names `oyb' and `fral'. The interaction is a failure if the listener's inventory does not contain the uttered word (here, `oyb'); in this case it is expanded to include it. Otherwise the interaction is a success, and both agents delete all alternative words (here, `fral' and `cherv') from their inventories.}
\end{figure}

The naming game is a model in which agents maintain an inventory of possible names for an object. In its first appearance in the physics literature \cite{bar06}, the dynamics were defined as follows. A speaker-listener pair is drawn from a population of $N$ agents at random, as in the Moran model. The speaker chooses one of the names in its inventory to present to the listener: if this inventory is empty, the speaker invents its own idiosyncratic name for the object and adds it to their inventory. The outcome of the interaction is deemed a success if the listener also has the uttered name in its inventory. In this case, any alternative names in both the speaker and listeners' inventories are removed. Otherwise the interaction is a failure and the listener adds the uttered name to the inventory.  These dynamics are illustrated in Figure~\ref{fig:namgam}.

Certain similarities with the Moran and utterance selection models are evident, for example the random sampling of agents from the population, and the random sampling of stored tokens (words) from the store (inventory) by the speaker. However, there are also a number of important differences. First, the inventory can accommodate an arbitrary number of words for the object, whereas in the utterance selection model the size of the store is fixed. Second, an innovation process is built into the dynamics from the outset, whereby agents invent a word for the object when they do not have one. Finally, the deletion rule introduces a bias into the replication process, whose nature and significance will become apparent as we discuss the effects of biases and selection mechanisms in a systematic way below. As with the utterance selection model, one can place speakers in the naming game on a social network, and vary the influence that certain individuals have over others, thereby implementing interactor selection (see e.g.,~\cite{bar06b,dal06}). It is perhaps worth noting that neither model described here implements replicator selection: all words or variants are considered to be equivalent in terms of their communicative function.

\subsection{Dynamics of consensus formation}

Armed with this pair of paradigmatic models of consensus formation, we now discuss how replication, innovation and selection interact to drive the emergence of shared linguistic behaviour within a speech community.

\subsubsection{Unbiased replication}

Even the simple stochastic replication process described by the Moran model, lacking innovation or any biases, is sufficient to achieve consensus. This can be understood from the stochastic equation of motion for this model, which can be derived fairly straightforwardly \cite{bly07}. First, let $x$ be the fraction of the $N$ agents in the speech community who use variant ${\rm A}$. In order for $x$ to change in one timestep, the speaker-listener pair must comprise one ${\rm A}$ and one ${\rm B}$ user. Then $x$ increases or decreases by $1/N$ with equal probability. Consequently, the expected change in $x$ is zero, and the stochastic process is purely diffusive. The rate of diffusion is obtained by examining the mean square change in $x$ per timestep: this is $2x(1-x)/N^2$. This allows one to write down the Fokker-Planck equation \cite{ris89}
\begin{equation}
\label{fpe}
\frac{\partial }{\partial t} P(x,t) = \frac{1}{\Lambda(N)} \frac{\partial^2}{\partial x^2} x(1-x) P(x,t) \;,
\end{equation}
where $P(x,t)$ is the probability that a fraction $x$ of the speech community are users of ${\rm A}$ at time $t$ given some initial condition, and the timescale $\Lambda(N) = N^2$ for the Moran model.  Formally, this equation is exact in the limit $N\to\infty$ under rescaling of time $\tau = t/\Lambda(N)$.

The equation (\ref{fpe}) is exactly solved for arbitrary initial condition \cite{kim55}, and therefore its dynamics are very well understood.  In particular it is known that eventually one of the two variants that are initially present will go extinct in a finite time. This is reminiscent of the Abrams and Strogatz' result for language competition, which showed that one of the competing languages is destined for extinction. Even without solving the equation (\ref{fpe}), it is clear that any timescale---including that of extinction---will be proportional to $\Lambda(N)$.

It turns out many variations on the basic Moran model, including the utterance selection model, are adequately described by (\ref{fpe}) with an appropriate definition of the variant frequency $x$ and characteristic timescale $\Lambda(N)$.   The earliest such model, introduced independently by Fisher \cite{fis30} and Wright \cite{wri31}, predates the Moran model by around 20 years, and describes changes in gene frequencies for species where generations do not overlap. Since in this case $N$ replication events occur concurrently, the characteristic timescale is $\Lambda(N)=2N$ (i.e., order $N$ shorter than in the Moran model, where replication events occur one-by-one). When individuals are placed on the sites of a square lattice, and speaker-listener pairs are restricted to neighbouring sites of the lattice, we obtain the voter model \cite{cli73} (although it was not actually referred to as such until a later work \cite{hol75}). When the number of spatial dimensions exceeds two, the Fokker-Planck equation adequately describes the dynamics of the total fraction of sites $x$ in state ${\rm A}$, and the characteristic timescale $\Lambda(N)\sim N^2$ as in the spatially unstructured case. In two dimensions, we also have (\ref{fpe}), but where now $\Lambda(N)$ is proportional to $N^2\ln(N)$.  On small-diameter networks, one yet again obtains (\ref{fpe}), but here the characteristic timescale $\Lambda(N)$ may scale with an exponent that depends on the network's degree distribution and the procedure for choosing a speaker and listener in each interaction \cite{soo05,soo08,bax08,bly10,bax12}.  In this case, it is also necessary to replace $x$ with a weighted average of a variant's frequency over the nodes of the network.  These general results apply both in the case where speakers can exhibit only categorical behaviour (as in the original Moran model) but also variable behaviour (as in the utterance selection model).

The relationship between the speech community size $N$ and the characteristic timescale $\Lambda(N)$ has proven crucial in evaluating descriptive theories of language change. For example, it was argued that a process of accommodation---whereby speakers seek to align their behaviour to each other in order to be better understood \cite{gil79}---was sufficient to explain the structure of new English language dialects \cite{tru04}. Whilst this is true of the final state of the language that is reached in the Moran and utterance selection models, the slow diffusion towards fixation is difficult to reconcile with the rapid rate at which such dialects are seen to form \cite{bax09}.

\subsubsection{Biased replication in language use}
\label{sec:usebias}

The Moran model demonstrates that replication alone can deliver consensus within a speech community in the absence of any other processes. However, it does so on a slow diffusive timescale. Biases in the dynamics, in production or perception, can serve to accelerate the consensus formation process in addition to a wide variety of other effects. One type of bias is a systematic bias in favour of one variant over another, that is replicator selection.  It is also possible for replication to be biased even when all variants are treated equally. The naming game provides one such example. Consider the successful interaction in Figure~\ref{fig:namgam}: the sole word remaining after the interaction is the one that was in the majority across the two speakers' inventories. Thus this model features a highly nonlinear \emph{regularisation bias} that boosts the frequency of a variant that has a local majority.

This type of bias, which is experimentally attested (as will be seen in Section~\ref{sec:acts} below) can be incorporated into the Moran model in a systematic way as follows.  The idea is to note that in the original Moran model, the probability, $f(x)$, that an agent using variant ${\rm A}$ is chosen as the speaker is equal to the frequency, $x$, of that variant. We can bias the selection of the speaker in favour of users of the majority variant by choosing $f(x) = x + \alpha b(x)$ in such a way that $b(x)>0$ when $x>\frac{1}{2}$ and $b(x)<0$ when $x<\frac{1}{2}$. If the variants are otherwise interchangeable, we must have $b(x)=-b(1-x)$. Moreover, if we do not allow re-introduction of extinct variants, we must furthermore have $b(0)=b(1)=0$. The lowest order polynomial with these properties is $b(x)=x(1-x)(2x-1)$.  In the regime where the bias is small, it contributes a drift term to the Fokker-Planck equation (\ref{fpe}) but leaves the diffusion term unchanged. The resulting Fokker-Planck equation can be written as a Langevin equation
\begin{equation}
\label{leone}
\dot{x}(t) = \alpha x(1-x)(2x-1) + \sqrt{x(1-x)} \eta(t)
\end{equation}
where $\eta(t)$ is a Gaussian white noise with zero mean and variance $1/\Lambda(N)$, and the multiplicative noise is interpreted in the Ito sense \cite{ris89,gar04}.  In this expression, $\alpha$ is the strength of the regularisation bias.

The idea of biasing the choice of speaker in this way is not particularly natural. However, similar equations arise when one considers the late-time regime of the naming game in which only two names remain. Agents in this model exhibit three states (knowledge of $A$, of $B$ and of both $A$ and $B$), and one finds a set of nonlinear differential equations governing their dynamics in the deterministic limit \cite{bar08}. A simple-minded dimensional reduction approach, which tracks the frequency of word $A$ across all inventories (whether alone or alongside $B$), yields to lowest order in $x$ an equation of the form (\ref{leone}) \cite{bly09}. The presence of this deterministic bias towards the majority variant leads to much faster extinction of the minority variant than when relying on the multiplicative noise alone. Numerical analysis of the full naming game dynamics \cite{bar06} shows that consensus is reached on a timescale of $O(N^{3/2})$ interactions. Moreover, on this timescale, the transition to the state of consensus becomes sharper as $N$ is increased, suggestive of something like a first-order dynamical phase transition.  As in the Moran model, the network topology can affect the rate of consensus formation. For example on $d$-dimensional lattices the ordering timescale scales as $N^{1+2/d}$ for $d\le4$ \cite{bar06b}, which for $d<2$ is the same ordering timescale as in the Moran model. For $d>2$, ordering in the naming game is faster, and on small-diameter networks one typically obtains the same behaviour as the in the case of $d>4$ and homogeneously-mixing populations, where the ordering timescale scales as $N^{3/2}$ interactions \cite{dal06}.

Even more subtle effects arise when a regularisation bias interact with other dynamical processes. An in-depth study of three-state models in the naming-game class, combined with population turnover that favours state $A$, can give rise to a discontinuous transition between consensus on state $A$, and a phase dominated by state $B$ \cite{col15}. This can be related to a sharp transition between the tendency for high-frequency verbs to be regular and low-frequency verbs to be irregular. (Note here that this regularisation process is distinct from that which serves to eliminate a minority variant, which is referred to as `regularisation of variation' by linguists).

Another example of a subtle interaction occurs when regularisation of variation is incorporated into the multi-speaker utterance selection model. Since a population in the Moran model becomes a speaker in the utterance selection model, the bias corresponds to speakers over-producing the variant that is in the majority in their memory.  If these speakers are placed on a square latter, we obtain the set of Langevin equations
\begin{eqnarray}
\label{lemany}
\dot{x}_i(t) &=&  \sum_{j} h_{ij} (x_j-x_i) +   \alpha x_i(1-x_i)(2x_1-1) \nonumber\\
&&\quad {} + \sqrt{x_i(1-x_i)} \eta_i(t)
\end{eqnarray}
where $x_i$ is the frequency of variant ${\rm A}$ in speaker $i$'s memory, $h_{ij}$ specifies the extent to which speaker $i$'s memory is affected by the tokens produced by their neighbours $j$ (see Figure~\ref{fig:moran}), and $\{\eta_i(t)\}$ is a set of independent Gaussian white noise terms. We can recognise the first term as a discretised Laplacian operator, and hence this term describes a spatial diffusion.

In this model, the strength of the Gaussian noise decreases as the size of the store $N$ is increased.  Thus for large $N$, one expects the noise to act as a weak perturbation on the deterministic part of the dynamics, which in turn is closely related to the Ginzburg-Landau model of coarsening in the Ising model with nonconserved order parameter. Indeed, Ising-like coarsening is observed at low noise strengths (at least on square lattices \cite{rus11}). However, for small $N$, the multiplicative nature of the noise acts as a strong perturbation on the dynamics to the extent that it in fact changes the universal properties of the coarsening dynamics from that of the Ising model to that of the voter model \cite{rus11}.  In more concrete terms, this means that when memory is short, speakers tend to exhibit categorical behaviour (because the strong noise tends to cause extinction at the level of individual speakers), and regions where speakers exhibit similar behaviour grow logarithmically in time. On the other hand, when memory is long, speakers exhibit variable behaviour and regions of similar behaviour grow as a power-law in time. 

Based on this result, we might speculate that for generic biases we expect there to be two regimes, one in which the noise serves to eliminate the bias and the phenomenology of the voter model is recovered, and one in which the noise is sufficiently weak that the deterministic limit of (\ref{lemany}) can be used to gain insights into the fate of variation under different types of bias. This latter regime was investigated systematically in \cite{bly12} by appealing to the different symmetry principles the biases may have. Asymmetry may arise separately in the coupling between speakers in (\ref{lemany}), that is when the parameters $h_{ij}$ in (\ref{lemany}) satisfy $h_{ij}\ne h_{ji}$, or in the form of the bias function $b(x)$, for example by taking $b(x)=x(1-x)$ which consistently favours variant ${\rm A}$.  These correspond to implementing interactor selection and replicator selection separately.  It is found that only the latter asymmetry provides a robust mechanism for reproducing the widely-observed S-curve pattern of language change, whereby the overall frequency of an innovation in the speech community tends to track a sigmoidal shape (such as a logistic function) as it propagates and establishes itself as a new convention \cite{bly12}. It is this kind of bias that can transform variation at the level of individual speakers, and translate it into a change in the macroscopic state of the language which was the fundamental dynamical process modelled at the topmost level of the hierarchy (see Section~\ref{sec:lineages}). However, this model of language change remains incomplete. For the propagation mechanism to work, the majority of speakers in the speech community must consistently favour one of the variants (e.g., ${\rm A})$ over the other.  Why this should be the case is difficult to explain without more detailed understanding of the origin and the nature of biases on linguistic behaviour, which we discuss in more detail in Section~\ref{sec:acts} below.

Another---perhaps more fundamental---puzzle with a model like that defined by (\ref{lemany}) is how to achieve stability of multiple dialects within a single speech community. This is related to the problem of obtaining stable coexistence between two languages discussed in Section~\ref{sec:communities}, and the reasons are in fact somewhat similar.  Suppose that the bias in (\ref{lemany}) is such that for some group of neighbouring individuals, the state of consensus (where all $x_i=1$ or all $x_i=0$ within this group) is stable. Then, if biases are uniform across the speech community, the state of global consensus will be stable. On way to achieve (at least metastable) coexistence of both variants across different members of the speech community is if the biases vary from one place to another, i.e., if one subset of individuals systematically boosts the frequency of one variant while another subset systematically boosts the frequency of the other variant. This is analogous to one of the proposed mechanisms for achieving coexistence of different languages.  However, such an explanation simply moves the goalposts: although we no longer have to explain where different language behaviour comes from, we must now explain how (much less tangible) differences in attitudes towards language behaviour come from, and how these become to be correlated with different regions in space. Studies in the area of opinion dynamics \cite{cas09} suggest one possible resolution of this difficulty in a self-consistent way, which is to incorporate a feedback whereby agents preferentially interact with those agents who behave in a similar way to themselves. Separation into groups that exhibit their own distinctive behaviour is seen in three classes of model that implement this idea: Brownian agents that diffuse towards regions of similar opinion \cite{sch00}, Deffuant-type models of bounded confidence \cite{lor07}, in which agents whose opinions differ by more than a certain amount have no effect on each other, and the Axelrod model \cite{axe97} and derivatives, in which agents exhibit multiple variable traits and the probability a pair of agents interacts depends on how many traits they have in common. In the language of Croft \textit{el al} \cite{cro00,bax09,bly12}, this corresponds to a feedback between replicator and interactor selection.

\subsubsection{Biased replication in language learning}
\label{sec:learnbias}

In the foregoing, we have mostly considered biases that apply at instances of language use. The effects of biases in language learning have been the focus of a considerable body of modelling work in the linguistics community, most notably within the framework of the iterated learning model \cite{kir02,smi14}. In these models, agents are typically characterised in terms of a \emph{grammar}---that is, a set of rules that governs the sentences they can produce. 

The task of an individual in the iterated learning model is to infer which of a set of grammars to adopt, given the sentences produced by other (typically more experienced) members of a speech community. As in the Moran model, the simplest case is to consider two grammars, ${\rm A}$ and ${\rm B}$. The dynamics of the iterated learning model are usually formulated as follows. A na\"ive agent is exposed to the utterances generated by an experienced agent's grammar, and infers a grammar (${\rm A}$ or ${\rm B}$) based on these utterances. At this point, they become an experienced agent and then serve as a model for one (or more) na\"ive agents to learn from. We can think of this as a replication process, in that grammars are replicated from one generation to the next: however, changes may occur in replication due to the fact that more than one grammar may be compatible with the set of utterances that the learner was exposed to. Biases may be present in production or learning that result in certain grammars being favoured over others as they are replicated: these thus mirror the biases discussed above in the context of language use and the utterance selection model. Indeed, a more formal connection between the iterated learning model and the utterance selection model was made in the case where a grammar specifies the frequency with which two variant forms should be used.  Here it was shown that if agents employ a Bayesian learning algorithm after hearing $N$ tokens of a variable structure, the estimated frequency $x$ is governed by the Langevin equation (\ref{leone}) with a bias that can act either in favour or against categorical use of a single variant, depending on the prior expectations of the language learner and the number of tokens heard \cite{rea09wal}.

The main purpose of iterated learning models is to understand processes by which an initially unstructured language may acquire structure after many generations of learning. Of particular interest are properties that are thought to transcend all languages. These \emph{design features} \cite{hoc60} include the fact language is productive (i.e., the ability to articulate a novel thought and be understood by a listener), exhibits duality of patterning (i.e., that combinatorial structure exists at both the level of meaningless and meaningful units) and compositional (i.e., that the meaning of an utterance can be understood from the meanings of the component parts). It seems likely that these two features are related: for example, compositionality provides a means by which language can be productive. What is less obvious is that a pressure on speakers to be productive can promote the emergence of compositional structure.

This effect was demonstrated within an iterated learning model with the space of grammars defined in such a way that agents could express a set of structured meanings (e.g., objects that have a shape and a colour) in either a holistic way (i.e., with no relation between the structure of the signal and meaning) or a compositional way (where the component parts of the meaning can be understood from the component parts of the signal) \cite{smi03}. The bias responsible for the emergence of structure lies in the production part of the interaction. The experienced agents have no control over which meanings they are required to express when providing training data to na\"ive agents: thus if the amount of data that is presented at each generation is small, agents may be confronted with meanings for which they have never heard a signal. Under these conditions, it is found that grammars that allow components of signals to be recombined are favoured by iterated learning. In more abstract terms, one might think of this as a competition between minimising the number of distinct signals in the grammar and maximising the ability to convey arbitrarily complex meanings. Such a competition may perhaps be related to guiding principles, such as the principle of least effort \cite{zip49} and the principle of uniform information density \cite{lev07}, which could potentially be investigated within this modelling framework. Meanwhile, it is worth noting that the grammars that feature in these models remain very simple, far away from the complexity of full-blown languages.

\subsubsection{Innovation}
\label{sec:ng}

The modelling of replication dynamics is greatly simplified if one knows in advance all possible variants of linguistic structure that may come into existence as the system evolves. Then, the state space is of fixed dimension, and one can (for example) extend the Moran model and its relatives to a larger number variants. Indeed, the multi-variant generalisation of (\ref{fpe}) is exactly solvable if the rate at which an agent spontaneously innovates a given variant $A_i$ is independent of the state of the system \cite{bax07}. Technically, this is because there is no current in the steady state: at some level, the dynamics fall into the same class as that of physical systems at thermal equilibrium. All other models of innovation generate nonequilibrium steady states, and we are not aware of any exact solutions for the multi-variant Moran model in this case.

The situation is even more complex when arbitrarily many variants can be created, as is the case for language, given its design feature of productivity. If one augments the Moran model with an innovation process that generates completely new variants at a constant rate, one arrives at the infinite-alleles model in population genetics for which some exact results are available \cite{ewe04,bly11}. By contrast, the full stochastic dynamics of models like the naming game, in which innovation is incorporated in a concrete and natural way, are resistant to analytical solution, and have therefore mostly been studied by direct Monte Carlo simulation.

From an empirical perspective, the most interesting application of a model in this class is to understand universal properties of colour terms. Languages differ in the number of basic colour terms (like `red' or `green' in English) that exist, and also in the shades that are regarded prototypical of a specific colour term \cite{ber69}. Nevertheless, the division of the continuous colour space into discrete categories is not completely arbitrary: in particular, the prototypes for each category (eg red, blue and green) form clusters in colour space that do not strongly overlap \cite{ber69,kay03}. Baronchelli \textit{et al} \cite{bar10} used an extension to the naming game to explain this phenomenon. The basic idea is that speakers are presented with multiple colours in a single scene, and seek to refer to a specific target colour. If the speaker's categorisation of the (one-dimensional) colour space is such that the target is the sole member of its category in the scene, a word associated with this category is presented to the listener. Otherwise, the speaker creates a new category (and a new word for it) that is sufficient to disambiguate the target from the other colours present. The interaction is successful if the listener recognises the word, and manages to correctly infer the target meaning (either because it is unambiguous according to the listener's categorisation, or because they guessed correctly among equally plausible alternatives). Addition and deletion of words than proceeds as in the naming game: in the case of failed communication, the speaker indicates the target meaning and the listener adds the uttered word to the category containing that meaning. Na\"ively one might expect these dynamics to lead to a very fine partitioning of colour space so that any combination of stimuli can be disambiguated. However, this is not the case for two reasons. First, only a small number of colours are typically shown in a scene, so the distance between them will typically be large. Second, speakers and listeners need to have well-aligned category boundaries for communication to be successful. Such alignment is hard to achieve with a large number of boundaries, and the failed interactions will tend to lead to words spreading between categories. Consequently, there is a trade-off between the ability to express distinctions between different colours and the difficulty of aligning a highly complex category system, reminiscent of the trade-offs seen in iterated learning models discussed above. The question of how category prototypes (which can be defined as the midpoint between two category boundaries) cluster around universal points was investigated by incorporating the psychological notion of a \emph{just noticeable difference} (JND) into the model. In the context of colour, this means that any two stimuli presented in the same scene differ in wavelength by an amount that exceeds some small value (otherwise the agent would not be able to distinguish them). Empirical research shows that the JND varies with wavelength: some nearby colours are easier for humans in all cultures to distinguish than others \cite{lon06}. Consequently, this breaks the translational symmetry in colour space, and generates clustering that is argued to be consistent with that found empirically \cite{bar10}. Models of this type have also been used to demonstrate the emergence of language design features (recall Section~\ref{sec:learnbias}), such as duality of patterning \cite{tri12}.

\section{Individual linguistic interactions}
\label{sec:acts}

Much of the modelling work described in the previous section is of a theoretical nature, aimed at understanding how processes of different types acting in the individual level shape the process of arriving at a shared linguistic conventions. In some cases contact with empirical data has been possible, for example the emergence of colour terms \cite{bar10}, new-dialect formation \cite{bax09} and verb regularisation \cite{col15}. It is however the case that historical data for language changes and other factors that might affect the process of change, such as social network structures, is relatively sparse (in comparison, for example, to data for financial transactions), thereby limiting the number of opportunities to apply models to real instances of language change. On the other hand, the last few years has seen a rapid growth in experimental research on human communicative and linguistic interactions (see \cite{sco10} for a recent review). Although these laboratory-based experiments are necessarily limited to small numbers of participants and short timescales, they provide valuable information about biases affecting individual linguistic behaviour. They also also illustrate some of the considerable complexity that is evident in human communication.

With this in mind, it is first worth considering how a communicative act would appear to alien with limited communicative capabilities (or perhaps a domestic cat).  What they would observe is people using their lungs and mouths to create pressure waves in the air, perhaps also moving their arms around at the same time, along with other movements like blinking, scratching of heads and so on.  What would it take for the alien to realise that these actions have a communicative intent?  This question has been addressed through experiments involving robots \cite{qui01} and humans \cite{sco09} that demonstrate processes by which communication channels are created on-the-fly through interactions without prespecifying that communication is desirable or providing an obvious way to do so. In the case of humans, the crucial step that the agents make is to perform behaviour that is unexpected from other agents' perspectives. This allows these agents to infer that the behaviour may be communicative \cite{sco09}. This process requires complex cognitive abilities---for example the ability to predict how another agent will respond to an action---and as such is argued to be specific to humans \cite{sco12}.

Given this understanding of how agents are able to identify certain behaviours as potential signals, the question now arises as to how they would associate a signal with a specific meaning. Here it is worth remarking that in almost all models discussed above in Section~\ref{sec:individuals}, the meaning of a signal was always assumed to be known to both agents in a communicative interaction. For example, in the case of the colour terms study \cite{bar10}, any ambiguity in the meaning of a word was always resolved after an interaction as the speaker ``unveils the topic [target meaning] via pointing'' \cite[SI:p2]{bar10}. If it is always possible to communicate a word's meaning by nonlinguistic means, then language is surely redundant.  This suggests that there is likely always to be some some ambiguity in any utterance, particularly in the process of language acquisition where children somehow infer the meanings of words (and more generally grammatical constructions) from experienced language users.

There is a considerable body of experimental research that addresses this problem (see e.g.~\cite{blo98} for a review of early contributions in this area).  Much of this work is conducted in an artificial language learning paradigm, whereby experimental participants are presented with a series of scenes containing a number of potential referents of a word (or set of words) and are asked questions that determine the inferences that participants have drawn about the meanings of words they have heard. The working hypothesis that has emerged is that a number of processes combine to deliver reliable word learning. First, on hearing an unfamiliar word, learners apply various \emph{heuristics} to  narrow down the (potentially infinite) set of possible meanings to a more manageable number of candidate meanings \cite{blo98}. Experimentally attested heuristics include following the gaze of the speaker \cite{bal91}, an expectation that a word is more likely to relate to a whole object rather than one of its parts \cite{lan88}, or that no two words have the same meaning \cite{mar88}. In some cases, these heuristics are able to eliminate all uncertainty and the word is learnt immediately. This effect is known as fast mapping \cite{car78}. However, it seems plausible that this is not always possible, and that language learners integrate information from multiple exposures to resolve any ambiguity. For example, if a child first encounters the word ``sheep'' on a visit to a farm, they could assume from the context that this refers to a cow that is also physically present; it is only on a later encounter that the cow meaning is implausible, whereas the sheep meaning is highly plausible, and the child revises their internal lexicon accordingly.

The general practice of combining information from multiple exposures is referred to as \emph{cross-situational learning} \cite{sis96}, and the separate process of hypothesising a (possibly incorrect) meaning and awaiting confirmation from later exposures is referred to variably as a \emph{guess-and-test} \cite{smi11} or \emph{propose-and-verify} \cite{tru13} heuristic. Both processes---along with a variety of others---been observed in artificial language learning experiments, which have been conducted variously with children and adults \cite{smi11,smi08,tru13}.  These results have inspired mathematical models of word learning which show that a lexicon of the size of that typically acquired by an adult human (60,000 words \cite{blo98}) can be learnt on a realistic timescale under quite general conditions \cite{bly10xsl,til12}. It is even in principle possible to learn the meaning of a word when an infinite set of alternative meanings can be inferred at any given exposure, as long as a learner can assign a plausibility ranking in such a way that the $k^{\rm th}$ most plausible meaning decays as a power-law $k^{-\sigma}$ with some positive exponent $\sigma$ \cite{bly14}. The main challenge faced by a cross-situational learner is when a nontarget meaning of a word is almost always as plausible as the target meaning. In this situation, the much-discussed mutual exclusivity constraint \cite{mar88}, whereby no two words are expected to have a common meaning, can greatly facilitate the process of identifying the correct meaning (as long as the plausible but incorrect meaning has an associated word) \cite{rei13}.

A small but significant (and somewhat ingenious) extension to the artificial language learning paradigm allows the emergence of structure that was previously seen in computational implementations of the iterated learning model (see Section~\ref{sec:learnbias}) to be replicated with real human participants. The idea is that the first participant learns a language that exhibits no systematic structure: for example, a random combination of syllables for a set of objects that occupy a structured meaning space (e.g., have various shape and colour combinations). After training, the participant is asked to name a set of objects (some of which they may not have seen before). The twist is that the answers that are provided in testing are then used as the training data for the next participant. This process can then be iterated to see how the language evolves as a consequence of repeated learning and production. The first experiment of this type \cite{kir08} confirmed the predictions of the earlier computer simulations \cite{smi03}, namely that the communication system acquires a systematic compositional structure over multiple generations of learning. This experiment further provides evidence for a competition between maximising the number of distinctions between meanings and minimising the complexity of the language that was alluded to above. This was seen by comparing two conditions, one in which participants could be exposed to a language where the same signal had multiple meanings, and one in which a filter was installed to prevent this from occurring. If the former case, the language became simpler by becoming less expressive; in the latter, the language retained its expressivity by becoming more structured.

Variations on this experimental approach have been used to explore other topics that were discussed in Sections~\ref{sec:individuals}.  Although inspired by experiments with robotic agents, the naming game has also been realised with human participants \cite{cen15}. Here the interest is in the role of the network topology that specifies which pairs of participants may interact, a subject that has been of great interest in the statistical physics community (see Section~\ref{sec:individuals} and \cite{cas09}).  This experiment demonstrated that consensus formation is accelerated in homogeneously mixing populations relative to spatial or random networks, as predicted by theoretical models \cite{bar06b,dal06}.

Other experiments have examined the role that biases may play in changing the way in which the frequency of different variants change over time. One set of experiments centres around a graphical communication task reminiscent of the game Pictionary \cite{gar07}, and has certain similarities to the naming game. Here, participants are provided with a predefined set of meanings and are asked to draw a picture to communicate one of them to another participant who has to identify the correct meaning. The set of meanings is deliberately constructed to contain items that are hard to differentiate graphically (e.g., `drama' and `soap opera'). In one such experiment \cite{gar07}, pairs of participants work through the set of meanings, thereby constructing their own individual symbols for the means. The participants are then grouped into new pairs, and asked to communicate the same set of meanings. A key question is what happens when two players who have established different symbols for the same meaning meet. Tamariz \emph{et al} \cite{tam14} consider two biases that might affect whether a participant keeps their existing symbol or switches to the other player's symbol after an interaction. The first is a coordination bias, which is a bias towards or against adopting symbols produced by other players. The second is a content bias, which is a bias towards specific symbols (e.g., because they communicated the intended meaning more effectively than others). These two biases are instances of interactor and replicator selection in the language of \cite{bax09,bly12}. The main findings were that players tended to be biased in favour of keeping their existing signals, but that evidence of content bias was present in most cases. It is interesting to note that these two effects are in conflict with each other: content bias promotes change, but agents are generally resistant to modifying their behaviour.

In the graphical communication experiment, agents are well characterised by usage of one particular signals for a given meaning. Since language users can be variable in language use, it is of interest to learn what happens to this variability when these users interact. In particular, this may provide some information as to the appropriate bias to include in an equation like Eq.~(\ref{leone}) which models variable language behaviour. Again, iterated learning experiments can be applied to this problem. Reali and Griffiths \cite{rea09} set up a language where each object has two names which appear with prescribed frequencies against their target meaning. After training, a participant is asked to name each object several times, which provides a set of frequencies to use for the next generation of learning.  In this approach, one can build up a transition matrix $T(x'|x)$, which gives the probability that when word $A$ is used with frequency $x$ to name an object, that after learning an individual uses that word with frequency $x'$. Reali and Griffiths find that in one generation of learning, learners appear to probability match, i.e., that $\bar{x}' = \sum_{x'} T(x'|x) \approx x$. However, considerable variation is evident, and over multiple generations of learning the frequency distribution that results is consistent with a weak bias against variability (i.e., one that favours frequencies close to $0$ or $1$). This is consistent with sociolinguistic understanding of variation, which is that where variation exists, it is usually conditioned on a linguistic factor (e.g., the pronunciation of a sound may depend on the position in the word) or a social factor (e.g., the speaker's age, level of education or other aspect of their identity) \cite{ogr11,mey11}. A further iterated learning experiment \cite{smi10} bears this out, wherein the initial language had consistent names for each of the objects, but two different plural markers. Although initially one plural marker is more frequent than the other, there is no correlation between which plural marking is used for which object. Here it is found that iterated learning either eliminates one of the plural markers, or that the variation becomes predictable: that is, each object is typically pluralised by one or other of the markers, but not both.

A final application of the artificial language learning paradigm is to investigate whether the nonuniform distribution of a structural feature over the world's languages (like word order, see Section~\ref{sec:lineages}) predicts the existence of a bias towards the more frequent structures in single instances of learning. One such study relates to affixes that change the meaning of the word. Where this occurs, it turns out that suffixes are more common than prefixes or infixes \cite{haw88}. To investigate whether this is also true at the individual level, St Clair \emph{et al} \cite{stc09} trained participants on an artificial language in which words are divided into two categories, each with their own affix. When presented with unfamiliar sentences, participants could more reliably identify when the correct affix was used for a particular word stem, in correspondence with the observation that suffixes are more common than prefixes.

As noted in Section~\ref{sec:lineages}, correlations between two structural features of a language have been observed, and are formulated as implicational universals. Culbertson \emph{et al} \cite{cul12} set up an experiment to determine whether a correlation between the order of numerals and nouns and the order of adjectives and nouns is visible at the level of individual learning events. In particular, one of the four possible orderings is much less frequent than the other three.  Participants were trained on four variant languages, and then asked to describe a set of scenes using the language they had learnt.  Points were awarded for a valid description of the scene (i.e., the right words, but in any order) and additionally if the order matched that of a computer-generated `native' speaker of the artificial language.  The input languages were set up in such a way that participants would score maximum points by using one of the four possible orderings exclusively.  The results of this experiment showed that this happened for all four input languages, apart from the one that corresponds to the ordering that is rare across the world's languages.  Again, this suggests that the nonuniform distribution over structures in the worlds' languages originates in biases that act at the level of single instances of language acquisition or use.

\section{Unifying and extending the hierarchy}
\label{sec:unify}

In this Colloquium I have surveyed four levels in a hierarchy of scales along a social dimension that is relevant to language dynamics.  One way to classify these levels is according to the main unit of variation at that level. This then determines the range of phenomena that are available for investigation: see Table~\ref{tab:hierarchy}.  Here, we have taken individual episodes of language acquisition and production to sit at the base of the hierarchy. These episodes typically involve a small number of language users (often just two) interacting for a short period of time (typically minutes). These length- and timescales are sufficiently small that laboratory experiments involving human participants---and the full complexity of their cognitive apparatus---can be used to measure biases that govern how language learning and production affects the structure and variability of language (see Section~\ref{sec:acts}).  With knowledge of these biases, we can then build effective models of individual human behaviour, and embed these individuals in a speech community, as described in Section~\ref{sec:individuals}. Through repeated interactions, language structure may emerge, and one that can characterise an entire speech community. At higher levels of the hierarchy, both speech communities and languages are taken as coherent units. In the former case this leads to competition between languages, as described in Section~\ref{sec:communities}, and in the latter, languages differentiate over time giving rise to a tree-like structure of relationships between them (Section~\ref{sec:lineages}).

The choice as to where to begin and end the hierarchy is entirely arbitrary, and an artifact of a constraint on the overall length of this article. At the bottom of the hierarchy, one could, for example, delve deeper into the workings of the human brain, and take the various cognitive capacities associated with language as the unit of variation. At the other end, we could extend to biological timescales, whereby these cognitive capacities are subject to biological evolution, and there is potential for an interaction between the cultural and biological evolutionary processes. 

\begin{table}
\begin{center}
\begin{tabular}{p{3.8cm}p{3.8cm}}
\hline\noalign{\smallskip}
Unit of variation & Locus of enquiry  \\
\noalign{\smallskip}\hline\noalign{\smallskip}
Language a shared system & Phylogenetic relationships between languages \\[0.75ex]
Speech community associated with a language & Language competition, birth, death and coexistence  \\[0.75ex]
Individuals within a speech community & Emergence of shared conventions and maintenance of variation between subcommunities \\[0.75ex]
Episodes of acquisition and production & Biases that favour one linguistic variant over another \\
\noalign{\smallskip}\hline
\end{tabular}
\end{center}
\caption{\label{tab:hierarchy} Hierarchy of scales in language dynamics along the social dimension covered in this article ordered by decreasing length- and timescales}
\end{table}

In principle, one could integrate the different levels of the hierarchy into a single whole, and the discussion we have given provides some detail as to how this might be achieved in practice. First, from the experiments described in Section~\ref{sec:acts}, we described how one can build up a transition matrix $T(x'|x)$ that specifies how individuals modify the frequency $x$ that they exhibit a specific behaviour in single interactions. This transition rate matrix could then be incorporated into the update rules of a Moran-type model (Section~\ref{sec:moran}), which would lead to a set of stochastic equations of motion analogous to Eq.~(\ref{lemany}), where the deterministic biases can be traced back to the experimentally-determined transition matrix $T(x'|x)$. From here, one can predict the structures that become established as conventions in a speech community, and by further augmenting individuals in the model with their own birth-death dynamics we can study the competition between languages with different structures that occurs at the level of the speech community (Section~\ref{sec:communities}). If one coarse-grains over individual variability and looks for a deterministic limit, it seems likely that a description similar to that postulated by Abrams and Strogatz, Eq.~(\ref{as}), will result. If one further coarse-grains over the number of speakers, and looks at the conventions that are formed over longer timescales, then an effective dynamics for changes in these structures---for example, the Poisson process (\ref{poisson}) that has been used in the phylogenetic analyses of Section~\ref{sec:lineages}---would result.

Although it is technically possible to connect the levels of the hierarchy in this way, is there any intellectual value in doing so? After all, when studying the relationships that emerge between languages at the largest length- and timescales, one may not be too interested in the precise process by which a new convention becomes established, just how this may be adequately described at the relevant timescales. A Poisson process could provide the adequacy that is desired.  However, in addressing this question it is worth looking more deeply into the some of the underlying motivations for research questions that have been posed at opposite ends of the hierarchy. For example, in their study of word-order universals at the level of entire languages, Maurits and Griffiths \cite{mau14} make reference to underlying psychological biases that act in favour of one word order (e.g., burdens on working memory) and how one would then expect these to be reflected in a preferred direction of change at the macroscopic level. Meanwhile, Culbertson \textit{et al} \cite{cul12} in their experimental study of changes in word order that occur at the level of an individual refer to the hypothesis that ``if a logically-possible grammatical system is not found, or is quite rate cross-linguistically, the explanation offered by [theories based on biases in learning] is that this system violates a learning bias'' \cite[p306]{cul12}.  It is reassuring that the results of both investigations point in a similar direction, in that less common word orders appear to be less stable both at the level of individual learning and at the level of historical change. However, to demonstrate that these are manifestations of the same phenomenon requires us to connect the two descriptions via the intermediate levels of the hierarchy. If this is achieved, one would gain deeper insights into which aspects of human cognition are most relevant to the process of cultural evolution, which in turn could help us understand how these cognitive abilities arose in the course of human evolution.

Two sources of complexity make such a unification challenging. First, many factors can affect language dynamics at each level of the hierarchy. By way of illustration, suppose that transmission matrices between linguistic states for single individuals have been established in an experimental setting, and are understood to characterise all human beings. On plugging these into speech-community-level models, it is clear from Section~\ref{sec:individuals} that changes in state at the level of the common language that emerges can depend on specific (nonlinguistic) features of those communities. For example, the characteristic timescales in the Moran model (denoted by the function $\Lambda(N)$ in Eq.~(\ref{fpe})) and in the naming game---and therefore likely most other models---vary with the speech community size $N$ in a way that depends on how the social network is connected, and the relative influence that different speakers can have on each other. At the same time, the propagation of a minority variant seems to require a positive disposition (replicator selection bias) towards it on the part of a majority of speakers \cite{bly12}. Such disposition towards a particular way of speaking presumably varies from one speech community to another and within a single community over time, for example due to its association with a particular well-regarded group of individuals \cite{cro00}. Thus it would in fact be somewhat surprising if the transition rates that apply at the macroscopic level, as in Eq.~(\ref{poisson}), were universal for a given structural feature in space and time, unless perhaps all the changes that occur at the lower levels of the hierarchy on such a fast timescale that they give the appearance of being so. Indeed, some studies \cite{lup10,dun11,bro15} provide evidence that language structure and the rate of language change covary with culture-specific factors.

The second source of complexity is one that we have barely touched on in this work, namely that human language is a highly complex object, with many different components (sounds, words, grammar), which presumably interact with each other in acquisition and use. Many models, empirical analyses and experiments on the \emph{dynamics} of language tend to employ fairly simple notions of language structure, for example, use of one of a number of variants of a single linguistic feature at the level of the language (as in Section~\ref{sec:lineages}) or individual speakers (as in Sections~\ref{sec:communities}--\ref{sec:acts}).  By contrast, investigations of the properties of a language (or across languages) at a single point in time are somewhat more sophisticated, and demonstrate the existence of a wide variety of statistical regularities, such as power-law word frequency distributions and long-range correlations in written texts \cite{zan14}, or universal properties of the semantic relationships between words in different languages \cite{you15}. At present, the origin of such phenomena remain to be understood within an emergentist framework: for example, the relative contributions from human cognition and social interactions in shaping these structures is not yet clear. Lying beyond such properties are more general design features, such as the ability for language to satisfy the productive needs of a speaker whilst remaining intelligible to a listener, which can only be understood if one moves outside frameworks in which the set of meanings and signals that can be communicated, and the structural relationships that exist between their components, are not artificially limited.

In summary, future advances in understanding language dynamics requires simultaneous progress on at least three fronts. First, a more complete mathematical understanding of what specific models predict would be valuable. For example, the exact solutions of the simplest Moran models and their variants have provided a great deal of insight into evolutionary dynamics in general \cite{ewe04}; however, these solutions are known only for the case where currents vanish in the steady state. Thus, as in physics, the more general case of athermal (entropy-producing) dynamics is the one that is both of greater applicability, but presents a correspondingly more difficult mathematical challenge. Meanwhile, more complex dynamical processes are typically implemented as computational models. Again, it is typically difficult to access realistic degrees of complexity (e.g., in speech community size, or structure of the linguistic system) in cases where the stationary state is not an equilibrium (zero-current) case, as the most efficient simulation algorithms tend to rely on this property.

Second, as this and other reviews (e.g.,~\cite{cas09,smi14}) attest, there is a profusion of models of language dynamics (and related processes like opinion formation). At the same time, many distinct models show broadly similar results, for example, a tendency towards consensus unless there is a pre-existing spatial symmetry breaking or a feedback between agent behaviour and the propensity to interact; or that structure tends to emerge over repeated learning. The statistical physics community in particular is yet to converge on a core set of models that achieve the optimal balance between (analytical or computational) tractability and explanatory power. It is also important to consolidate the various modelling approaches that have been taken into a set of general principles. A number of works hint for example at principles that involve trade-offs between the expressive power and complexity of a language system, which are broadly reminiscent of the trade-offs between energy and entropy that provide a deep understanding of collective phenomena in physical systems.

Finally, it is a truth almost universally acknowledged that data for language behaviour over the course of human history is scarce. Although the World Atlas of Language Structures \cite{wals} is extensive in terms of the number of languages and structures that it covers, and augmented by companion databases such as the Atlas of Pidgin and Creole Language Structures \cite{apics}, these document languages at only one point in time. This allows historical changes to be inferred, but validating these inferences by comparing with the actual changes that have occurred is difficult. These difficulties are further confounded if it turns out that additional factors (such as the speech community size or social network structure) also affect historical language dynamics, since then these will need to be known as well. This highlights the need for good-quality data sets that track language use and social structures over time. An obvious source of such data, if one is specifically interested in how linguistic behaviour spreads and changes over short timescales, may be derived from social media. In particular, it may be possible to determine whether models of the type described in Section~\ref{sec:individuals} adequately describe how innovations propagate through an online community. Other large-scale sources of data with greater time depth also exist, such as Google's corpus of scanned books \cite{goo}. Alongside providing snapshots of language use, the online world also offers the opportunity for mass behavioural experiments. Many of experimental designs outlined in Section~\ref{sec:acts} lend themselves to being set as tasks on platforms like the Amazon Mechanical Turk, and their results have been found to correlate well with traditional experiments (see e.g.~\cite{sch10}). Scientific progress is greatly accelerated when data is abundant, and with some creative thinking, data-driven advances in language modelling may pave the way to a more complete understanding of processes that lie beyond the grasp of empirical research.

\begin{acknowledgement}
I think Bill Croft for introducing me to this fascinating topic, and Simon Kirby, Andrew Smith and Kenny Smith for their ability to explain linguistics in a way that a physicist can understand.
\end{acknowledgement}

%
\bibliographystyle{epj}
\bibliography{micmacs}

\end{document}